\documentclass[a4paper,nofootinbib,twocolumn]{revtex4}

\usepackage[T1]{fontenc} 
\usepackage{graphicx}

\begin{document}

\title{Attractive Hubbard Model: Homogeneous Ginzburg -- Landau Expansion
and Disorder}

\author{$^1$E.Z. Kuchinskii, $^1$N.A. Kuleeva, $^1$$^,$$^2$M.V. Sadovskii}

%\footnote{E-mail: kuchinsk@iep.uran.ru, strigina@iep.uran.ru, sadovski@iep.uran.ru}

\affiliation{$^1$Institute for Electrophysics, Russian Academy of Sciences, 
Ural Branch,\\
Amundsen str. 106, Ekaterinburg 620016, Russia\\
$^2$M.N. Mikheev Institute for Metal Physics, Russian Academy of Sciences, Ural Branch,\\
S. Kovalevskaya str. 18, Ekaterinburg 620290, Russia}

%\date{\today}

\begin{abstract}

We derive Ginzburg -- Landau (GL) expansion in disordered attractive Hubbard 
model within the combined Nozieres -- Schmitt-Rink and DMFT+$\Sigma$
approximation. Restricting ourselves to the case of homogeneous
expansion, we analyze disorder dependence of GL expansion coefficients
on disorder for the wide range of attractive potentials $U$, from weak
BCS coupling region to the strong coupling limit, where superconductivity
is described by Bose -- Einstein condensation (BEC) of preformed Cooper
pairs. We show, that for the case of semi -- elliptic ``bare'' density of
states of conduction band, disorder influence on GL coefficients
$A$ and $B$ before quadratic and fourth -- order terms of the order
parameter, as well as on the specific heat discontinuity at superconducting
transition, is of universal nature at any strength of attractive interaction and
is related only to the general widening of the conduction band by disorder.
In general, disorder growth increases the values of coefficients $A$ and $B$,
leading either to the suppression of specific heat discontinuity (in the weak
coupling limit), or to its significant growth (in the strong coupling region).
However, this behavior actually confirms the validity of the generalized 
Anderson theorem, as disorder dependence of superconducting critical temperature 
$T_c$, is also controlled only by disorder widening of conduction band
(density of states).
\end{abstract}

\maketitle

\section{Introduction}

The problem of superconductivity in BCS --- BEC crossover region (and up to the strong 
coupling limit) has a long history, starting with early works by Leggett and Nozieres 
and Schmitt-Rink [\onlinecite{Leggett,NS}].
Probably the simplest model to study this crossover is Hubbard model with attractive
interaction. The most successive approach to the studies of Hubbard model (both repulsive
and attractive) is the dynamical mean field theory (DMFT) 
[\onlinecite{pruschke,georges96,Vollh10}]. Attractive Hubbard model was already studied 
within DMFT in a number of papers [\onlinecite{Keller01,Toschi04,Bauer09,Koga11,JETP14}]. 
However, up to now there are only few works, where disorder effects were taken into
account, either in normal or superconducting phases of this model. Qualitative analysis
of disorder effects upon critical temperature $T_c$ in BCS --- BEC crossover region was
presented in Ref. [\onlinecite{PosSad}], which claimed the validity of Anderson theorem
in this region for the case of $s$-wave pairing. Diagrammatic analysis of disorder 
effects on $T_c$ and the properties of the normal state in crossover region was
recently presented in Ref. [\onlinecite{PalStr}].

We have developed the generalized DMFT+$\Sigma$ approach to Hubbard model
[\onlinecite{JTL05,PRB05,FNT06,UFN12}], which is quite convenient for the account of
different ``external'' interactions, e.g. such as disorder scattering 
[\onlinecite{HubDis,HubDis2}]. This approach is also well suited to the studies of
two--particle properties, such as dynamic (optical) conductivity  [\onlinecite{HubDis,PRB07}].  
In a recent paper [\onlinecite{JETP14}] we used this approach to analyze the single--particle
properties of the normal phase and optical conductivity in attractive Hubbard model.
Further on the DMFT+$\Sigma$ approximation was combined with Nozieres -- Schmitt-Rink  
approach  to study the influence of disorder on superconducting critical 
temperature $T_c$ in BCS -- BEC crossover and strong coupling region
[\onlinecite{JTL14,JETP15}], demonstrating the validity of the generalized
Anderson theorem. Disorder effects upon $T_c$ are essentially due only the general
widening of the conduction band by random scattering. This was demonstrated exactly 
(for the whole range of attractive interactions) in the case of semi -- elliptic density 
of states of conduction band (three-dimensional case) at any disorder level and becomes
also valid in the case of flat band (two-dimensional case) in the limit of strong 
enough disorder.

Ginzburg -- Landau (GL) expansion in the region of BCS -- BEC crossover was derived in
a number of previous papers [\onlinecite{Micnas01,Zwerger92,Zwerger97}], however no
effects of disorder scattering on GL -- expansion coefficients was considered.
Here we derive the microscopic coefficients of (homogeneous) GL -- expansion for the 
attractive Hubbard model and study disorder effects on these coefficients including 
the BCS -- BEC and strong coupling regions, as well as upon the specific heat discontinuity 
at superconducting transition, demonstrating certain universality of disorder behavior of
these characteristics..

\section{Disordered Hubbard model in DMFT+$\Sigma$ approach}

We consider the disordered attractive Hubbard model with Hamiltonian:
\begin{equation}
H=-t\sum_{\langle ij\rangle \sigma }a_{i\sigma }^{\dagger }a_{j\sigma
}+\sum_{i\sigma }\epsilon _{i}n_{i\sigma }-U\sum_{i}n_{i\uparrow
}n_{i\downarrow },  
\label{And_Hubb}
\end{equation}
where $t>0$ is transfer integral between the nearest neighbors and 
$U$ is onsite Hubbard attraction , 
$n_{i\sigma }=a_{i\sigma }^{\dagger }a_{i\sigma }^{{\phantom{\dagger}}}$ 
is electron number operator at site $i$, $a_{i\sigma }$ ($a_{i\sigma }^{\dagger}$) 
is annihilation (creation) operator of an electron with spin $\sigma$. 
Local energy levels $\epsilon _{i}$ are assumed to be independent random 
variables on different lattice sites. We assume the Gaussian distribution 
of $\epsilon _{i}$ at each site for the validity of the 
standard ``impurity'' scattering diagram technique [\onlinecite{Diagr}]:
\begin{equation}
\mathcal{P}(\epsilon _{i})=\frac{1}{\sqrt{2\pi}\Delta}\exp\left(
-\frac{\epsilon_{i}^2}{2\Delta^2}
\right)
\label{Gauss}
\end{equation}
Here $\Delta$ is the measure of disorder scattering. 

The generalized DMFT+$\Sigma$ approach [\onlinecite{JTL05,PRB05,FNT06,UFN12}] supplies the
standard DMFT [\onlinecite{pruschke,georges96,Vollh10}] with an additional ``external'' 
self-energy (in general case momentum dependent) due to any interaction outside the DMFT,
which provides an effective method to calculate both single and two -- particle properties 
[\onlinecite{HubDis,PRB07}]. The additive form of the total self-energy conserves the
structure of self -- consistent equations of DMFT [\onlinecite{pruschke,georges96,Vollh10}]. 
The ``external'' self-energy is recalculated at each step of the standard DMFT iteration
scheme, using some approximations, corresponding to the form of an additional interaction,
while the local Green's function (central for DMFT) is also ``dressed'' by additional
interaction.

For disordered Hubbard model we take the ``external'' self-energy entering DMFT+$\Sigma$ 
cycle in the simplest form of self -- consistent Born approximation, neglecting the
``crossing'' diagrams due to disorder scattering:
\begin{equation}
\tilde\Sigma(\varepsilon)=\Delta^2\sum_{\bf p}G(\varepsilon,{\bf p}),
\label{BornSigma}
\end{equation}
where $G(\varepsilon,{\bf p})$ is the complete single -- particle Green's function. 

To solve the effective Anderson impurity model of DMFT here we used the effective
algorithm of numerical renormalization group (NRG) [\onlinecite{NRGrev}].

In the following, we consider the model of the ``bare'' conduction band with semi -- elliptic
density of states (per unit cell and spin projection):
\begin{equation}
N_0(\varepsilon)=\frac{2}{\pi D^2}\sqrt{D^2-\varepsilon^2}
\label{DOSd3}
\end{equation}
where $D$ determines the half -- width of conduction band. This is a good approximation for
the three -- dimensional case.

In Ref. [\onlinecite{JETP15}] we have shown analytically, that in DMFT+$\Sigma$ approach,
within these approximations, all the disorder influence upon single -- particle properties
is reduced to the simple effect of band widening by disorder scattering, so that 
$D\to D_{eff}$, where $D_{eff}$ is the effective band half -- width in the presence of
disorder (in the absence of correlations, i.e. for $U=0$):
\begin{equation}
D_{eff}=D\sqrt{1+4\frac{\Delta^2}{D^2}}.
\label{Deff}
\end{equation}
and conduction band density of states (in the absence of $U$) ``dressed'' by disorder is
given by:
\begin{equation}
\tilde N_{0}(\varepsilon)=\frac{2}{\pi D_{eff}^2}\sqrt{D_{eff}^2-\varepsilon^2}
\label{tildeDOS}
\end{equation}
conserving its semi -- elliptic form. 

For other models of the  ``bare'' conduction band density of states, besides band widening,
disorder scattering changes the form of the density of states, so that the complete universality
of disorder influence of single -- particle properties, strictly speaking, is absent. However,
in the limit of strong enough disorder the ``bare'' band density effectively becomes elliptic
for any reasonable model, so that universality is restored [\onlinecite{JETP15}].

All calculation below were performed for the quarter -- filled band (n=0.5 per lattice site).

\section{Ginzburg -- Landau expansion}

The critical temperature of superconducting transition $T_c$ in attractive
Hubbard model was analyzed using direct DMFT calculations a number of papers 
[\onlinecite{Keller01,Toschi04,Koga11}]. In Ref. [\onlinecite{JETP14}] we have
determined $T_c$ from instability condition of the normal phase (instability
of DMFT iteration procedure). The results obtained in this way were in good
agreement with the results of Refs. [\onlinecite{Keller01,Toschi04,Koga11}].
Additionally, in Ref. [\onlinecite{JETP14}] we calculated $T_c$ using the
approximate Nozieres -- Schmitt--Rink approach in combination with DMFT (used
to calculate the chemical potential of the system), demonstrating that being
much less time consuming, it provides semi -- quantitative description $T_c$ 
behavior in BCS -- BEC crossover region, in good agreement with direct 
DMFT calculations. In Refs. [\onlinecite{JTL14,JETP15}] the
combined Nozieres -- Schmitt-Rink approach was used to study the detailed
dependence of $T_c$ on disorder. Below we shall use this combined approach
to derive GL -- expansion including the disorder dependence of GL -- expansion
coefficients.

\begin{figure}
\includegraphics[clip=true,width=0.48\textwidth]{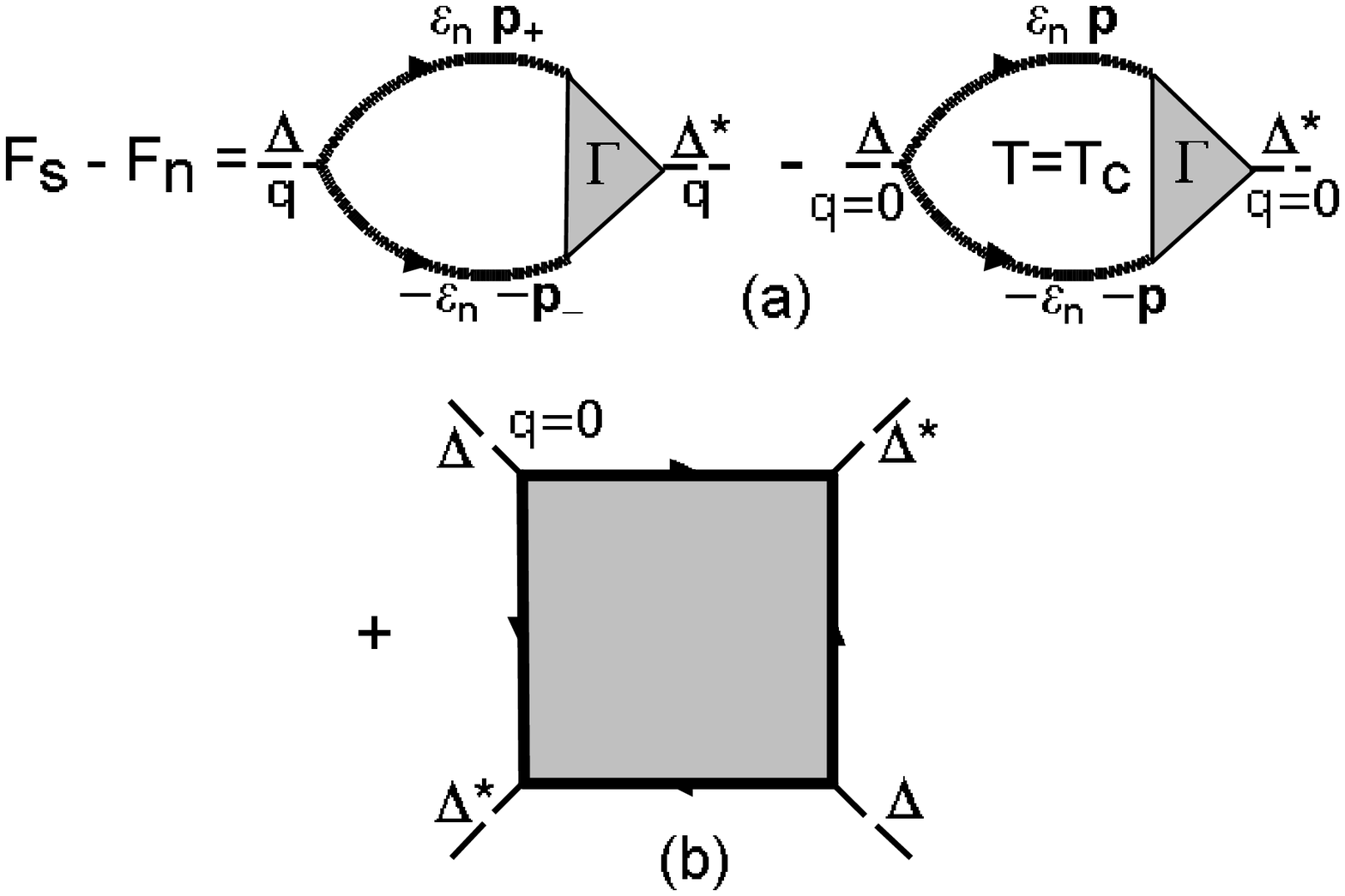}
\caption{Diagrammatic representation of Ginzburg -- Landau expansion.}
\label{diagGL}
\end{figure}

We shall write GL -- expansion for the difference of free energies of 
superconducting and normal phases in the standard form:
\begin{equation}
F_{s}-F_{n}=A|\Delta_{\bf q}|^2
+q^2 C|\Delta_{\bf q}|^2+\frac{B}{2}|\Delta_{\bf q}|^4,
\label{GL}
\end{equation}
where $\Delta_{\bf q}$ is the spatial Fourier component of the amplitude of 
superconducting order parameter. Microscopically, this expansion  is determined 
by diagrams of the loop -- expansion for the free energy of an electron in the 
``external field'' of (static) superconducting order parameter fluctuations 
with small wave vector ${\bf q}$ , shown in Fig.\ref{diagGL} (where fluctuations 
are represented by dashed lines) [\onlinecite{Diagr}].
Below we limit ourselves to the case of homogeneous expansion with ${\bf q}=0$
and calculations of its coefficients $A$ and $B$, leaving the 
(much more complicated) analysis of the general inhomogeneous case of finite 
${\bf q}$ and calculations of coefficient $C$ in (\ref{GL}) for the future work.

Within Nozieres -- Schmitt-Rink approach [\onlinecite{NS}] we use the weak 
coupling approximation to calculate loop -- diagrams with two and four Cooper
vertices shown in Fig. \ref{diagGL}, dropping all corrections due to Hubbard
$U$, while including ``dressing'' by disorder scattering\footnote{In the 
absence of disorder this approach just coincides with that used in
Refs. [\onlinecite{Micnas01,Zwerger92,Zwerger97}], using Hubbard --
Stratonovich transformation in the functional integral over fluctuations of
superconducting order parameter.}. However, the chemical potential, which
essentially depends on the coupling strength $U$ and determines the condition of
BEC in the strong coupling region, is calculated via the full DMFT+$\Sigma$ 
procedure.

Coefficient $A$ before the square of the order parameter in GL -- expansion
is given by  diagrams of Fig. \ref{diagGL}(a) with $q=0$ [\onlinecite{Diagr}]:
\begin{equation}
A(T)=\chi_0(q=0,T)-\chi_0(q=0,T_c),
\label{A_det}
\end{equation}
where
\begin{equation}
\chi_0(q=0,T)=-T\sum_{n}\sum_{\bf pp'}\Phi_{\bf pp'}(\varepsilon_n)
\label{chi}
\end{equation}
is the two -- particle loop in Cooper channel ``dressed'' only by disorder
scattering, while $\Phi_{\bf pp'}(\varepsilon_n)$ is disorder averaged
two -- particle Green's function in Cooper channel ($\varepsilon_n=\pi T(2n+1)$
is corresponding Matsubara frequency). Subtraction of the second diagram
in Fig. \ref{diagGL}(a), i.e. that of $\chi_0(q=0,T_c)$ in (\ref{A_det}),
guarantees the validity of $A(T=T_c)=0$, which is necessarily so in any kind
of Landau expansion [\onlinecite{Diagr}].

To obtain $\sum_{\bf pp'}\Phi_{\bf pp'}(\varepsilon_n)$ we use an exact Ward 
identity, derived in Ref. [\onlinecite{PRB07}]:
\begin{eqnarray}
G(\varepsilon_n,{\bf p})-G(-\varepsilon_n,-{\bf p})=\nonumber\\
=-\sum_{\bf p'}\Phi_{\bf pp'}(\varepsilon_n)(G_0^{-1}(\varepsilon_n,{\bf p'})-G_0^{-1}(-\varepsilon_n,-{\bf p'})),
\label{Word}
\end{eqnarray}
Here $G(\varepsilon_n,{\bf p})$ is disorder averaged (but not ``dressed'' by
Hubbard interaction!) single -- particle Green's function.
Using the symmetry $\varepsilon({\bf p})=\varepsilon(-{\bf p})$ and
$G(\varepsilon_n,-{\bf p})=G(\varepsilon_n,{\bf p})$, we obtain from  the Ward
identity (\ref{Word}):
\begin{equation}
\sum_{\bf pp'}\Phi_{\bf pp'}(\varepsilon_n)=
-\frac{\sum_{\bf p}G(\varepsilon_n,{\bf p})-\sum_{\bf p}G(-\varepsilon_n,{\bf p})}{2i\varepsilon_n},
\label{Phi}
\end{equation}
so that for Cooper susceptibility (\ref{chi}) we get:
\begin{eqnarray}
&&\chi_0(q=0,T)=\nonumber\\
&&=T\sum_{n}\frac{\sum_{\bf p}G(\varepsilon_n,{\bf p})-\sum_{\bf p}G(-\varepsilon_n,{\bf p})}{2i\varepsilon_n}=\nonumber\\
&&=T\sum_{n}\frac{\sum_{\bf p}G(\varepsilon_n,{\bf p})}{i\varepsilon_n}.
\label{chi1}
\end{eqnarray}
Performing now the standard summation over Matsubara frequencies 
[\onlinecite{Diagr}], we obtain:
\begin{eqnarray}
&&\chi_0(q=0,T)=\nonumber\\
&&=\frac{1}{4\pi i}\int_{-\infty}^{\infty}d\varepsilon
\frac{\sum_{\bf p}G^R(\varepsilon,{\bf p})-\sum_{\bf p}G^A(\varepsilon,{\bf p})}{\varepsilon}th\frac{\varepsilon}{2T}=\nonumber\\
&&=-\int_{-\infty}^{\infty}d\varepsilon\frac{\tilde N(\varepsilon)}{2\varepsilon}th\frac{\varepsilon}{2T},
\label{chi2}
\end{eqnarray}
where $\tilde N(\varepsilon)$ is the ``bare'' ($U=0$) density of states,
``dressed'' by disorder scattering, which in the case of semi -- elliptic
band takes the form (\ref{tildeDOS}).  
In Eq. (\ref{chi2}) the origin of $\varepsilon$ is at the chemical potential.
Replacing $\varepsilon\to\varepsilon -\mu$ to move the origin of energy to the
center of conduction band, we finally write:
\begin{equation}
\chi_0(q=0,T)=-\int_{-\infty}^{\infty}d\varepsilon
\frac{\tilde N(\varepsilon)}{2(\varepsilon -\mu)}th\frac{\varepsilon -\mu}{2T},
\label{chi_end}
\end{equation}
Cooper instability of the normal phase, determining superconducting transition
temperature $T_c$, is written as:
\begin{equation}
1=-U\chi_0(q=0,T_c)
\label{Cupper}
\end{equation}
Then, to determine the critical temperature we obtain the following equation:
\begin{equation}
1=\frac{U}{2}\int_{-\infty}^{\infty}d\varepsilon \tilde N_0(\varepsilon)\frac{th\frac{\varepsilon -\mu}{2T_c}}{\varepsilon -\mu} ,
\label{BCS}
\end{equation}
Using (\ref{Cupper}) to determine $\chi_0(q=0,T_c)$ and (\ref{chi_end}) for 
$\chi_0(q=0,T)$, we obtain the coefficient $A$ (\ref{A_det}):
\begin{equation}
A(T)=\frac{1}{U}-
\int_{-\infty}^{\infty}d\varepsilon \tilde N_0(\varepsilon)
\frac{th\frac{\varepsilon -\mu }{2T}}{2(\varepsilon -\mu )}.
\label{A_end}
\end{equation}
The chemical potential for different values of $U$ and $\Delta$ is to be
determined here from direct DMFT+$\Sigma$  calculations, i.e. from the
standard equation for the total number of electrons (band filling), defined
by Green's function obtained in DMFT+$\Sigma$ approximation.
This allows us to find both $T_c$ and GL -- expansion coefficients in the
wide range of parameters of the model, including the BCS -- BEC crossover
region and the limit of strong coupling, for different disorder levels.
Actually, this is the essence of Nozieres -- Schmitt-Rink approximation ---
in the weak coupling region transition temperature is controlled by the
equation for Cooper instability, while in the strong coupling limit it is
defined as the temperature of Bose condensation, which is controlled by the
equation for chemical potential. The joint solution of Eqs. (\ref{BCS}) and
(\ref{A_end}) with DMFT+$\Sigma$ equation for chemical potential provides the 
correct interpolation for $T_c$ and GL -- coefficient $A$ from weak coupling 
region via the BCS -- BEC crossover towards the strong coupling. 

For $T\to T_c$ the coefficient $A(T)$ is written as:
\begin{equation}
A(T)\equiv a(T-T_c).
\label{A2}
\end{equation}
where in case of temperature independent chemical potential
\begin{equation}
a=\frac{1}{4T_c^2}\int_{-\infty}^{\infty}d\varepsilon
\tilde N_0(\varepsilon)\frac{1}{ch^2\frac{\varepsilon -\mu}{2T_c}}.
\label{AA2}
\end{equation}
In BCS approximation with conduction band of infinite width with constant 
density of states $\tilde N_0(0)$ we obtain from (\ref{AA2}) the standard  
result $a=\frac{\tilde N_0(0)}{T_c}$ [\onlinecite{Diagr}]. However, in
BCS -- BEC crossover region temperature dependence of $\mu$ is essential
and we have to use the general expression (\ref{A_end}) in conjunction
with equation for $\mu$ to calculate $a$. At the same time,
from Eq. (\ref{A_end}) it is clear that disorder scattering influences $a$
only through the changes of the density of states $\tilde N_0(\varepsilon)$ 
and chemical potential $\mu$, which is the typical single -- particle property.
Thus, in the case of semi -- elliptic ``bare'' conduction band the dependence
of $a$ on disorder is due only to the band widening by disorder replacing
$D\to D_{eff}$. Thus, in the presence of disorder we expect the universal
dependence of $a(2D_{eff})^2$ on $U/2D_{eff}$ (all energies are to be 
normalized by the effective bandwidth $2D_{eff}$), which will be confirmed by
the results of direct numerical computations in the next Section (cf. Fig.
\ref{fig3}(a)).

Coefficient $B$ is determined by ``square'' diagram with four Cooper vertices with
${\bf q}=0$, ``dressed'' in arbitrary way by disorder scattering, which is 
shown in Fig. \ref{diagGL}(b) [\onlinecite{Diagr}]:
\begin{eqnarray}
B=\frac{1}{2}T\sum_{n}\sum_{{\bf p}_1{\bf p}_2{\bf p}_3{\bf p}_4}
<G(i\varepsilon_n;{\bf p}_1,{\bf p}_2)G(-i\varepsilon_n;-{\bf p}_2,-{\bf p}_3)\nonumber\\
G(i\varepsilon_n;{\bf p}_3,{\bf p}_4)G(-i\varepsilon_n;-{\bf p}_4,-{\bf p}_1)>,\nonumber\\
\label{B_det}
\end{eqnarray}
where $<\cdots >$ denotes averaging over disorder, while 
$G(i\varepsilon_n;{\bf p}_1,{\bf p}_2)$ (and other similar expressions) 
represent exact single -- particle Green's functions for the fixed configuration 
of the random potential. Performing standard summation over Matsubara
frequencies, we obtain:
\begin{eqnarray}
B=\frac{1}{2}\int_{-\infty}^{\infty}\frac{d\varepsilon}{2\pi i}th\frac{\varepsilon}{2T}
\sum_{\bf p_1p_2p_3p_4}
<G^R(\varepsilon;{\bf p_1},{\bf p_2})\nonumber\\
G^A(-\varepsilon;-{\bf p_2},-{\bf p_3})
G^R(\varepsilon;{\bf p_3},{\bf p_4})G^A(-\varepsilon;-{\bf p_4},-{\bf p_1})>. 
\nonumber\\
\label{B1}
\end{eqnarray}
Due to zero value of momentum ${\bf q}=0$ in Cooper vertices and  the static 
nature of disorder scattering, we can now use certain generalization of the
Ward identity (\ref{Word}) to get (at $T=T_c$): 
\begin{equation}
B=\int_{-\infty}^{\infty}\frac{d\varepsilon}{4\varepsilon ^3}
\left(th\frac{\varepsilon}{2T_c}-\frac{\varepsilon/2T_c}{ch^2\frac{\varepsilon}{2T_c}}\right)
\tilde N_0(\varepsilon)
\label{B_end1}
\end{equation}
Detailed derivation is presented in Appendix \ref{app}. In BCS approximation,
using the conduction band of infinite width with constant density of states 
$\tilde N_0(0)$, we immediately obtain from Eq. (\ref{B_end1})
the standard result: $B=\frac{7\zeta(3)}{8\pi^2 T_c^2}\tilde N_0(0)$ 
[\onlinecite{Diagr}].

Again, replacing here $\varepsilon\to\varepsilon -\mu$, to move the origin of
energy to the middle of the conduction band, we can write:
\begin{equation}
B=\int_{-\infty}^{\infty}\frac{d\varepsilon}{4(\varepsilon -\mu)^3}
\left(th\frac{\varepsilon -\mu}{2T_c}-\frac{(\varepsilon -\mu)/2T_c}{ch^2\frac{\varepsilon -\mu}{2T_c}}\right)
\tilde N_0(\varepsilon)
\label{B_end}
\end{equation}
It is seen, that disorder dependence of the coefficient $B$ (similarly to $A$) 
is also determined only by disorder widened density of states $\tilde N_0(\varepsilon)$ 
and chemical potential, so that in the case of semi -- elliptic ``bare'' conduction 
band it is reduced to simple replacement $D\to D_{eff}$, leading to universal 
dependence of $B(2D_{eff})^3$ on $U/2D_{eff}$, which is confirmed by the results of 
direct numerical computations presented in the next Section and shown in Fig.\ref{fig3}b.

It should be stressed, that Eqs. (\ref{A_end}) and (\ref{B_end}) for
GL -- coefficients $A$ and $B$ were obtained with the use of exact Ward
identities, and are thus valid also in the limit of strong disorder
(beyond Anderson localization). 

Universal dependence on disorder, related to conduction band widening by
disorder scattering, is also valid for specific heat discontinuity at $T_c$,
as it is completely determined by coefficients $a$ and $B$:
\begin{equation}
C_s(T_c)-C_n(T_c)=T_c\frac{a^2}{B}
\label{Cs-Cn}
\end{equation}
Appropriate numerical results are also given in the next Section 
(cf. Fig. \ref{fig4}(b)).

Coefficient $C$ before the gradient term of GL -- expansion is determined
essentially by two -- particle characteristics (due in particular to 
non -- trivial $q$ -- dependence of the vertex, which is obviously
changed by disorder scattering). In particular, the behavior of $C$ is
significantly changed at Anderson transition [\onlinecite{SCLoc}], so that
no universality of disorder dependence is expected in this case. 

\section{Main results}

Let us now discuss the main results of our numerical calculations,
directly demonstrating the universal dependencies of GL -- coefficients
$A$ and $B$ and specific heat discontinuity at $T_c$ on disorder. 

\begin{figure}
\includegraphics[clip=true,width=0.45\textwidth]{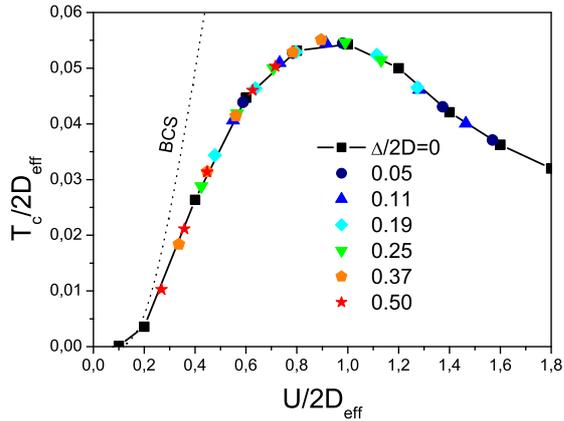}
\caption{Universal dependence of superconducting critical temperature on
disorder for different values of Hubbard attraction.}
\label{fig1}
\end{figure}

In Fig. \ref{fig1} we show the universal dependence of critical temperature
$T_c$ on Hubbard attraction $U$ for different levels of disorder,
which was obtained and discussed in detail in Refs. [\onlinecite{JTL14,JETP15}]. 
Typical maximum of $T_c$ at $U/2D_{eff}\sim 1$ is characteristic of BCS -- BEC 
crossover region.

\begin{figure}
\includegraphics[clip=true,width=0.45\textwidth]{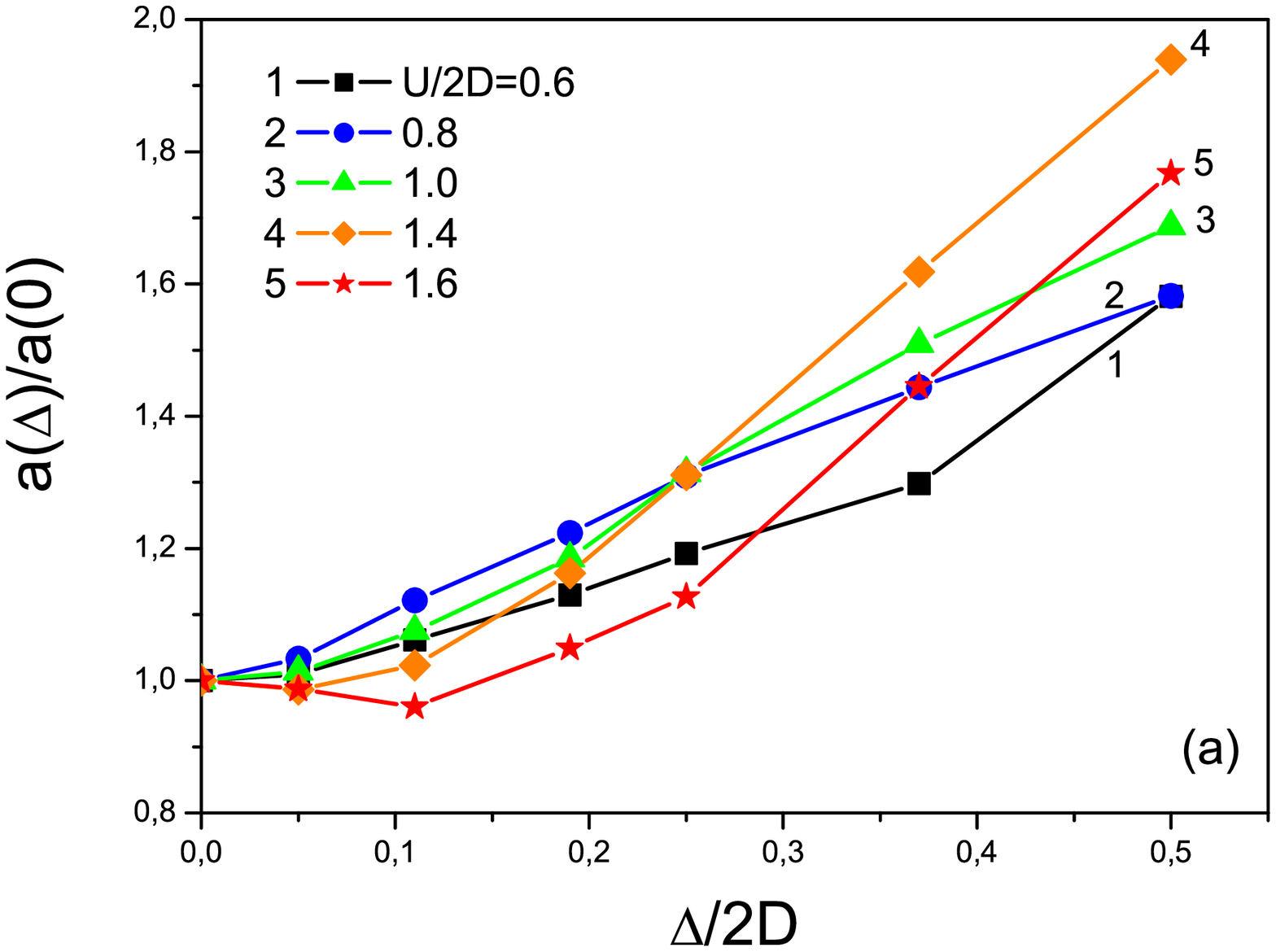}
\includegraphics[clip=true,width=0.45\textwidth]{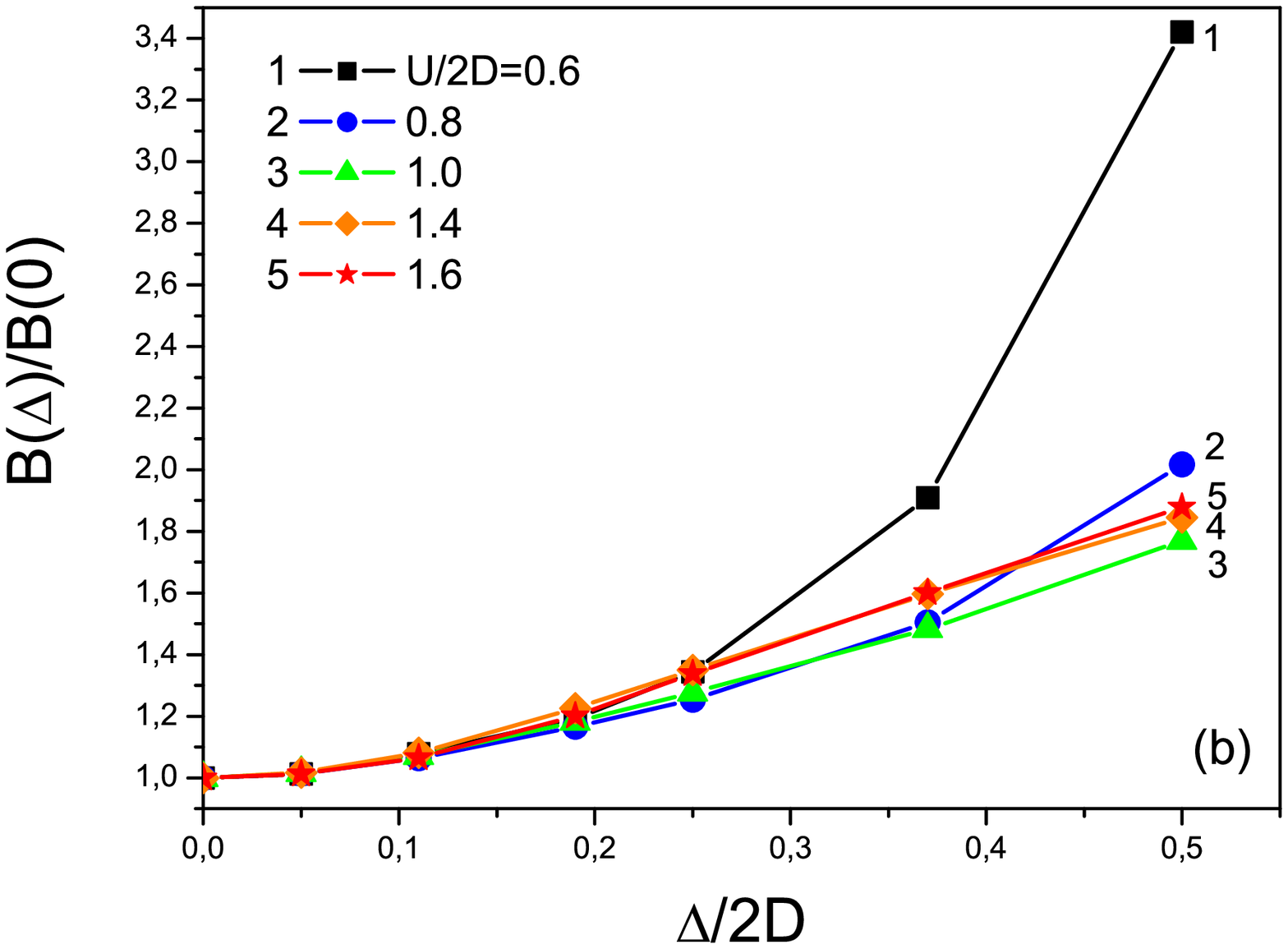}
\caption{Disorder dependence of GL -- coefficients $a$ (a) and 
$B$ (b), normalized by their values in the absence of disorder,
for different values of Hubbard attraction.}
\label{fig2}
\end{figure}

In Fig. \ref{fig2} we present disorder dependencies of GL -- coefficients
$a$ (Fig. \ref{fig2}(a)) and $B$ (Fig. \ref{fig2}(b)) for different values of
Hubbard attraction. We can see that $a$ in general increases with the growth of
disorder. Only in the limit of strong enough coupling $U/2D > 1.4$ 
(curves 4 and 5) in the region of weak disorder we observe weak suppression
of $a$ by disorder scattering. Coefficient $B$  pretty fast grows with disorder
in the region of weak coupling (curve 1 in Fig. \ref{fig2}(b)), while in the
region of strong coupling this growth becomes more moderate 
(curves 4,5 in Fig. \ref{fig2}(b), so that in this region the dependence of 
$B$ on disorder becomes almost independent of the value of $U$
(curves 4 and 5 practically coincide). 

\begin{figure}
\includegraphics[clip=true,width=0.45\textwidth]{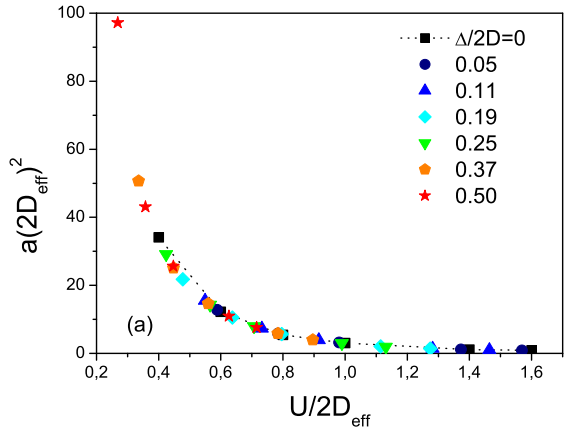}
\includegraphics[clip=true,width=0.45\textwidth]{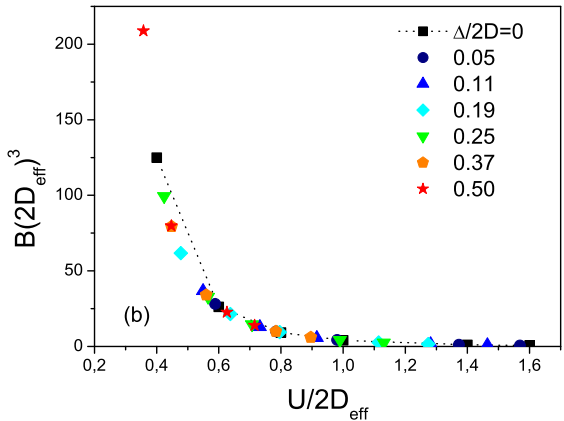}
\caption{
Universal dependence of GL -- coefficients $a$ (a) and $B$ (b) on
Hubbard attraction for different values of disorder.}
\label{fig3}
\end{figure}

However, this rather complicated dependence of coefficients
$a$ and $B$ on disorder is determined solely by the growth of effective 
conduction bandwidth with disordering given by Eq. (\ref{Deff}).
In Fig. \ref{fig3} we show the universal dependencies of GL -- coefficients
$a$ (a) and $B$ (b), normalized by appropriate powers of effective bandwidth,
on the strength of Hubbard attraction. In the absence of disorder (dashed
line with squares) coefficients $a$ and $B$ drop fast with the growth of $U$.
Other symbols in Fig. \ref{fig3} show the results of our calculations for
different levels of disorder. It is clearly seen, that all the data ideally fit
the universal curve, obtained in the absence of disorder.

\begin{figure}
\includegraphics[clip=true,width=0.45\textwidth]{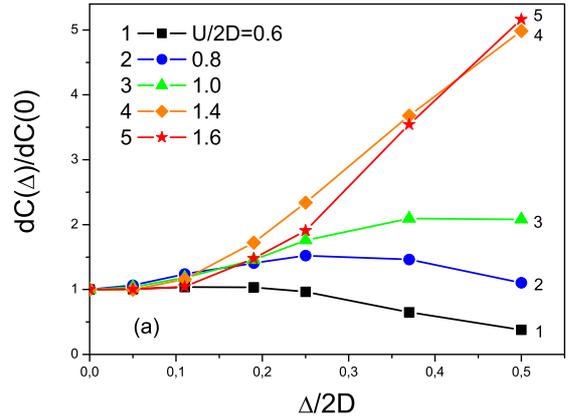}
\includegraphics[clip=true,width=0.45\textwidth]{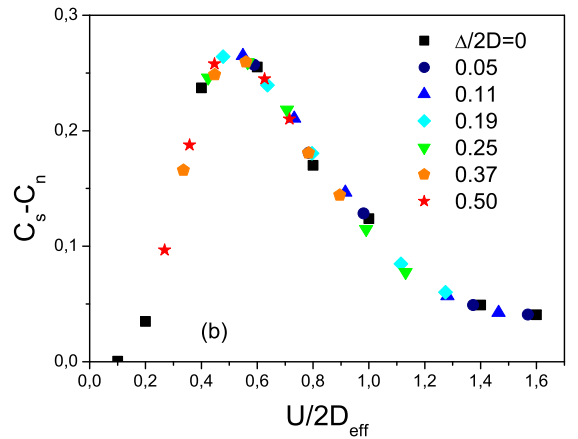}
\caption{Dependence of specific heat discontinuity at critical
temperature $dC\equiv C_{s}-C_{n}$ on disorder for different values
of Hubbard attraction $U$ (a) and universal dependence of this
discontinuity on  $U$ for different values of disorder (b).}
\label{fig4}
\end{figure}

Coefficients $a$ and $B$ determine specific heat discontinuity at the
critical temperature (\ref{Cs-Cn}). As these coefficients and $T_c$
[\onlinecite{JTL14,JETP15}] depend on disorder in universal way due only to
the growth of the effective bandwidth  (\ref{Deff}), the same type of
universal dependence is also valid for specific heat discontinuity.
In Fig. \ref{fig4}(a) we show the dependence of specific heat discontinuity
$dC\equiv C_{s}-C_{n}$ on disorder for different values of Hubbard attraction
$U$. It is seen, that in the region of weak coupling (curve 1) specific heat
discontinuity is suppressed by disordering, for intermediate couplings
(curves 2,3) weak disorder leads to the growth of specific heat discontinuity,
while the further growth of disorder suppresses this discontinuity.
In the region of strong coupling (curves 4,5) the growth of disorder leads to
significant growth of specific heat discontinuity, which is mainly related to
the similar growth of $T_c$ (cf. [\onlinecite{JTL14,JETP15}]). 
However, this complicated dependence of specific heat discontinuity on
disorder is again completely determined by the growth of the effective
bandwidth (\ref{Deff}). In Fig.\ref{fig4}(b) we show the universal dependence
of specific heat discontinuity on $U$, normalized by the bandwidth $2D_{eff}$.
Black squares represent data in the case of absence of disorder. Other symbols
in Fig. \ref{fig4}(b) show the data for different disorder levels.
It is seen again, that all the data precisely fit the universal dependence of
specific heat discontinuity obtained in the absence of disorder.
Specific heat discontinuity grows with the growth of $U$ in the region of
weak coupling $U/2D_{eff}\ll 1$ and drops with the growth of $U$ in the limit
of strong coupling $U/2D_{eff}\gg 1$. The maximum of specific heat discontinuity
is observed at $U/2D_{eff}\approx 0.55$. Actually, this dependence of specific
heat discontinuity qualitatively resembles the similar dependence of critical 
temperature, though the its maximum is reached at smaller values of Hubbard 
attraction.

\section{Conclusion}

Using the combination of Nozieres -- Schmitt-Rink approximation 
with the generalized DMFT+$\Sigma$ approach we have studied disorder 
influence upon coefficients $A$ and $B$ determining the homogeneous 
Ginzburg --- Landau expansion and specific heat discontinuity at 
superconducting transition in attractive Hubbard model. 

We have demonstrated analytically, that in the
case of the ``bare'' conduction band with semi -- elliptic density of
states disorder influence on GL -- coefficients $A$ and $B$ and specific
heat discontinuity is universal and is controlled only by the general
conduction band (density of states) widening by disorder scattering and
illustrated this conclusion with explicit numerical calculations,
performed for the wide range of attractive potentials $U$, from weak 
coupling region where $U/2D_{eff}\ll 1$ and superconducting instability 
is described by the usual BCS approach, up to the strong coupling region 
where $U/2D_{eff}\gg 1$ and superconducting transition is determined by 
Bose -- Einstein condensation of preformed Cooper pairs.

These results essentially prove the validity of the generalized Anderson 
theorem in BCS -- BEC crossover region and in the limit of strong coupling 
not only for superconducting $T_c$ [\onlinecite{JTL14,JETP15}], but also
for homogeneous Ginzburg -- Landau expansion, determining appropriate 
thermodynamic effects, like specific heat discontinuity at transition point.

This work is supported by RSF grant 14-12-00502.

%\newpage

\appendix

\section{Coefficient $B$ in the presence of disorder} 

\label{app}
Coefficient $B$ is determined by ``square'' diagram with four Cooper 
vertices with ${\bf q}=0$, ``dressed'' by disorder scattering, shown in 
Fig. \ref{diagGL}(b). 
Corresponding analytic expression was given above in Eq. (\ref{B_det}). 
After the standard summation over Matsubara frequencies $B$ is written as in 
(\ref{B1}), i.e. is determined by the following combination of four Green's 
functions with real frequencies:
\begin{eqnarray}
\sum_{\bf p_1p_2p_3p_4}
<G^R(\varepsilon;{\bf p_1},{\bf p_2})G^A(-\varepsilon;-{\bf p_2},-{\bf p_3})\nonumber\\
G^R(\varepsilon;{\bf p_3},{\bf p_4})G^A(-\varepsilon;-{\bf p_4},-{\bf p_1})>. 
\label{Ap1}
\end{eqnarray}
where $<\cdots >$ denotes averaging over disorder and 
$G^{R(A)}(\varepsilon;{\bf p}_1,{\bf p}_2)$ are the exact retarded (advanced) 
single -- particle Green's functions for the fixed configuration of disorder.

\begin{figure}
\includegraphics[clip=true,width=0.48\textwidth]{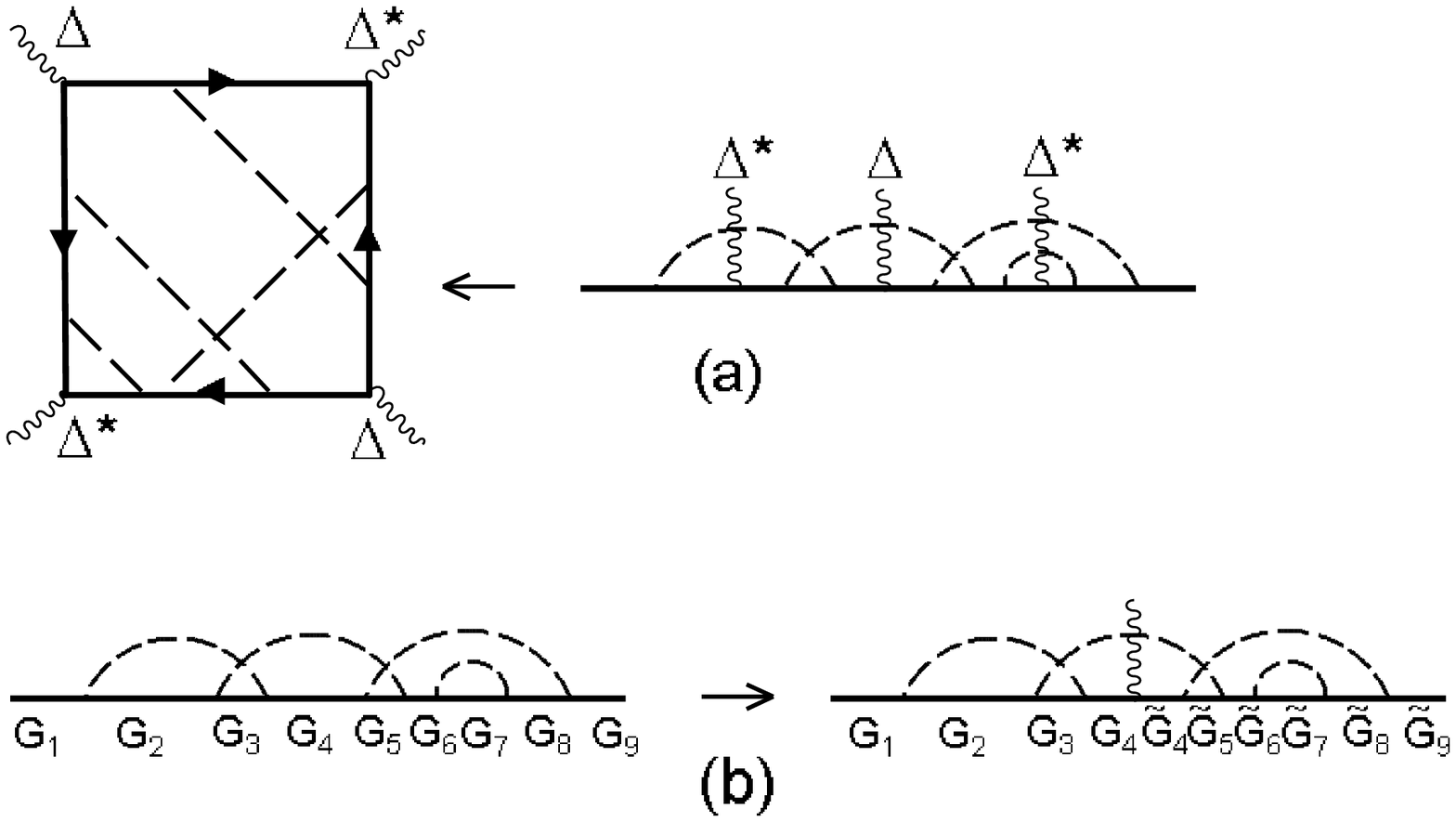}
\caption{Diagrams for coefficient $B$ and derivation of the
generalized Ward identity.}
\label{diagTW}
\end{figure}

Typical diagram for the fourth order of disorder scattering (dashed lines) is
shown in Fig. \ref{diagTW}(a). Arbitrary diagrams for such four -- particle
Green's function can be obtained from diagrams for single -- particle Green's
function of the same order of disorder scattering by arbitrary inserting 
three Cooper vertices into ``bare'' electron Green's functions, as shown in
Fig. \ref{diagTW}(a). Taking into account the static nature of disorder
scattering and zero value of transferred momentum ${\bf q}=0$ in Cooper
vertices, we can evaluate (\ref{Ap1}) using certain generalization of exact 
Ward identity (\ref{Word}), derived in Ref. [\onlinecite{PRB07}].

Let us take the diagram for single -- particle Green' function, shown in the
left part of Fig. \ref{diagTW}(b),  and consider certain configuration of
momenta transferred by dashed lines. Here we have nine ``bare'' electron
Green's functions with momenta ${\bf p}_1\cdots {\bf p}_9$. In the following
we use short notations:
\begin{equation}
G_i=G_0^R(\varepsilon;{\bf p_i})\qquad \tilde G_i=G_0^A(-\varepsilon;-{\bf p_i}),
\label{Ap2}
\end{equation}
where $G_0^{R(A)}(\varepsilon;{\bf p})=\frac{1}{\varepsilon-\varepsilon ({\bf p})\pm i\delta}$ 
is the ``bare'' Green's function. Insertion of Cooper vertex leads to the sign change 
of momenta and frequencies (i.e. to the replacement $G_i\leftrightarrow\tilde G_i$) 
in all Green's functions standing to the right from the vertex.
Let us assume, that the central of three Cooper vertices was inserted in the fourth
Green's function, as shown in the right part of Fig. \ref{diagTW}(b). 
Arbitrary insertion of the first Cooper vertex into one of the first four of
Green's functions leads to the following result:
\begin{eqnarray}
G_1G_2G_3G_4\to 
G_1\tilde G_1\tilde G_2\tilde G_3\tilde G_4+G_1G_2\tilde G_2
\tilde G_3\tilde G_4+\nonumber\\
G_1G_2G_3\tilde G_3\tilde G_4+G_1G_2G_3G_4
\tilde G_4,
\label{Ap3}
\end{eqnarray}
so that taking into account $G_i^{-1}-\tilde G_i^{-1}=2\varepsilon$, we get:
\begin{eqnarray}
G_1\tilde G_1\tilde G_2\tilde G_3\tilde G_4\frac{G_1^{-1}-\tilde G_1^{-1}}{2\varepsilon}+\cdots\nonumber\\
+G_1G_2G_3G_4\tilde G_4\frac{G_4^{-1}-\tilde G_4^{-1}}{2\varepsilon}=
\nonumber\\
=\frac{\tilde G_1\tilde G_2\tilde G_3\tilde G_4-G_1G_2G_3G_4}{2\varepsilon}
\label{Ap4}
\end{eqnarray}
Then $\tilde G_4\tilde G_5\tilde G_6\tilde G_7\tilde G_8\tilde G_9\to G_4G_5G_6G_7G_8G_9$ and
after all insertions of the last (third) Cooper vertex in one of the six Green's
functions  $G_4\cdots G_9$, we again obtain:
$\frac{\tilde G_4\tilde G_5\tilde G_6\tilde G_7\tilde G_8\tilde G_9-G_4G_5G_6G_7G_8G_9}{2\varepsilon}$.

Thus we get:
\begin{eqnarray}
&&<G^R(\varepsilon)G^A(-\varepsilon)G^R(\varepsilon)G^A(-\varepsilon)>=\nonumber\\
&&=<\frac{G^A(-\varepsilon)-G^R(\varepsilon)}{2\varepsilon}
\frac{G^A(-\varepsilon)-G^R(\varepsilon)}{2\varepsilon}>=
\nonumber\\
&&=\frac{1}{4\varepsilon^2}(<G^A(-\varepsilon)G^A(-\varepsilon)>+<G^R(\varepsilon)G^R(\varepsilon)>-\nonumber\\
&&-2<G^R(\varepsilon)G^A(-\varepsilon)>)=
\nonumber\\
&&=\frac{1}{4\varepsilon^2}
\left\{\frac{d}{d\varepsilon}(<G^A(-\varepsilon)>-<G^R(\varepsilon)>)-\right.
\nonumber\\
&&\left.-\frac{<G^A(-\varepsilon)>-<G^R(\varepsilon)>}{\varepsilon}\right\},
\label{Ward1}
\end{eqnarray}
where we can evaluate two -- particle Green's functions with $q=0$ again using the analogue 
of the Ward identity (\ref{Word}) for real frequencies.
Using (\ref{Ward1}) in (\ref{B1}) and making in terms with $<G^A(-\varepsilon)>$ 
under the integral over $\varepsilon$ the replacement $\varepsilon\to -\varepsilon$, 
we obtain the following expression for coefficient $B$: 
\begin{eqnarray}
&&B=\int_{-\infty}^{\infty}\frac{d\varepsilon}{2\pi i}
\frac{th\frac{\varepsilon}{2T}}{4\varepsilon^2}\times\nonumber\\
&&\times\left(\frac{d}{d\varepsilon}-\frac{1}{\varepsilon}\right)
(\sum_{\bf p}G^A(\varepsilon,{\bf p})-\sum_{\bf p}G^R(\varepsilon,{\bf p}))=
\nonumber\\
&&=\int_{-\infty}^{\infty}d\varepsilon\frac{th\frac{\varepsilon}{2T}}{4\varepsilon^2}
\left(\frac{d}{d\varepsilon}-\frac{1}{\varepsilon}\right)\tilde N_0(\varepsilon)=\nonumber\\
&&=\int_{-\infty}^{\infty}\frac{d\varepsilon}{4\varepsilon^3}
\left(th\frac{\varepsilon}{2T}-\frac{\varepsilon /2T}{ch^2\frac{\varepsilon}{2T}}\right)
\tilde N_0(\varepsilon)
\label{B2}
\end{eqnarray}
which was used in the main part of the paper.

%\newpage

\newpage

\end{document}